\documentstyle[preprint,aps,floats,epsf]{revtex}

\newcommand{\notE}{\ \hbox{{$E$}\kern-.60em\hbox{/}}}
\newcommand{\notp}{\ \hbox{{$p$}\kern-.43em\hbox{/}}}
\def\D0{\mbox{D\O}}

\begin{document}

\tightenlines


\preprint{
\font\fortssbx=cmssbx10 scaled \magstep2
\hbox to \hsize{
\hfill$\vcenter{
\hbox{\bf MADPH-99-1138}
\hbox{\bf hep-ph/9911510}}$ }
%
}

\title{
\vspace*{2cm}
Phenomenology of a String-Inspired Supersymmetric Model \\ 
with Inverted Scalar Mass Hierarchy}

\author{V. Barger, Chung Kao and Ren-Jie Zhang}

\address{Department of Physics, University of Wisconsin, 
1150 University Avenue, Madison, WI 53706}

\maketitle

\thispagestyle{empty}

\begin{abstract}

Supersymmetric (SUSY) models 
with heavy sfermions ($m_{\tilde{f}} \sim 10$ TeV) 
in the first two generations 
and the third generation sfermion masses below 1 TeV
can solve the SUSY flavor and the CP problems 
as well as satisfy naturalness constraints.
We study the phenomenology of a string-inspired scenario 
and compare it with the minimal supergravity unified model (mSUGRA).
The SUSY trilepton signature at the upgraded Tevatron, 
the $b\rightarrow s\gamma$ branching fraction  
and the neutralino dark matter relic density in this model 
can differ significantly from the mSUGRA model. 

\end{abstract}

\pacs{PACS numbers: 13.85.Qk, 13.90.+i, 14.80.Bn, 14.80.Ly}
%


\section{Introduction}

Generic supersymmetric (SUSY) extensions of the standard model (SM) 
may generate large flavor-changing neutral current (FCNC)
and CP violation effects. 
Many entries in the sfermion mass matrices and some CP-violating phases 
must be sufficiently suppressed to satisfy stringent experimental bounds,
e.g. from the $K-\bar{K}$ mass difference,
CP asymmetries in the kaon system, and 
the electric dipole moments of the electron 
and the neutron \cite{Gabbiani:1996hi}. 
These quantities have been assumed to be small 
in the minimal supergravity unified model (mSUGRA) 
without much justification.
Providing a satisfactory solution to the problem is 
one of the major motivations in SUSY model building 
\cite{Nir:1993mx,Dimopoulos:1995mi,Dvali:1996rj,Bagger:1999ty,Baer:1999md}.

Several models have been proposed to realize a scenario 
with heavy sparticles in the first two generations \cite{Dimopoulos:1995mi}.
In these models the heavy soft masses are of order 10 TeV 
to solve the SUSY flavor and CP problems, while 
the third generation sfermions and the Higgs bosons 
still have soft masses of weak scale order, thus satisfying
the naturalness condition constraints.
This inverted scalar mass hierarchy is well motivated 
because the most stringent constraints from FCNC 
and CP violation processes only apply to the first two generations.
(The first two generations can also be subject to naturalness constraints
through a one-loop $D$-term which however is zero in unification models.)
In Ref.~\cite{Dvali:1996rj},
the $D$-term of an anomalous U(1) symmetry 
(which is common to many 4-D string models) has been used to 
generate an inverted mass hierarchy--the $D$-term dominates over 
the gravity-mediated $F$-term contribution to the sfermion soft masses
if U(1) charges are appropriately assigned.
In Ref.~\cite{Bagger:1999ty,Baer:1999md},
the hierarchy is generated from grand-unified scale 
soft masses of order a few TeV 
through renormalization group (RG) evolution; the third generation 
sfermion and Higgs boson masses are highly suppressed at the weak scale
because of the associated large third generation Yukawa couplings 
and the infrared fixed points
of the RG equations. In the latter models, the gravitino mass is also 
of order of several TeV, solving the problem of late 
gravitino decay after the period of big bang nucleosynthesis
\cite{deCarlos:1993jw}.

In this Letter we first consider a string-inspired model that generates
an inverted scalar mass hierarchy. 
This model consists of 
the minimal supersymmetric standard model superfields and 
two singlet chiral superfields $S$ and $T$ with nonzero $F$-component
vacuum expectation values (VEVs). 
We assume $F_S \simeq 10^{-2} F_T\simeq M_W M_{pl}$, where $M_W$ is
the weak scale. 
Similar relations $F_S\ll F_{T}$ appear naturally 
in many models with gaugino condensation \cite{Lalak:1995hn}. 
The gaugino mass is determined from the gauge kinetic term, 
which is of the form 
\begin{equation}
\int d^2\theta~{S\over M_{pl}} W^a W_a\ .
\end{equation}
This gives $m_{1/2}\simeq F_S/M_{pl}\simeq M_W$. The sfermion masses
are determined from the K\"ahler potential, for which we take 
the following form 
\begin{equation}
K(S,T,Q_i)\ =\ -\log(S+{\bar S})-3\log(T+\bar{T}) 
          +(T+\bar{T})^{n_i} Q_i^\dagger Q_i,
\label{KP}
\end{equation}
where $n_i$ is the overall modular weight for the matter field $Q_i$,
and $i$ the family index. Choosing $n_3=-1$ and $n_{1,2}>-1$, we find
\begin{equation}
{\tilde m}_{1,2}, A_{1,2}\simeq {F_T\over M_{pl}}\simeq 100~M_W, \qquad
{\tilde m}_{3}, A_{3}\simeq {F_S\over M_{pl}}\simeq M_W,
\end{equation}
where $\tilde{m}_i$ is the $i$-th generation sfermion mass.
Therefore this choice of modular weights generates an inverted mass 
hierarchy at the string scale.
The gravitino mass in this model, $m_{3/2}\simeq F_T/M_{pl}\simeq 100~M_W$,
is of order of $10$ TeV. 
The form of the K\"ahler potential in Eq.~({\ref{KP})
can be obtained in 4-D heterotic string models; indeed, fields from the
untwisted sector have the modular weight $-1$, and our model corresponds
to the moduli-dominated SUSY breaking 
scenario of Ref. \cite{Brignole:1994dj}.

In the following, we study the phenomenology of 
this model with an inverted scalar mass hierarchy (the ISM model).  
We set the first-two-generation sfermion masses and trilinear couplings 
at 5 TeV, and assume a common scalar mass ($m_0$) 
and a common trilinear coupling ($A_0$) for the third generation sfermions, 
along with a universal gaugino mass ($m_{1/2}$)
at the grand unified scale ($M_{\rm GUT}$): 
\begin{eqnarray}
&& m_{1/2}, \;\; \tilde{m}_1 = \tilde{m}_2 = 5 \;\; {\rm TeV}, \;\; 
   \tilde{m}_3 = m_0, \nonumber \\
&& A_1 = A_2 = 5 \;\; {\rm TeV}, \;\;  A_3 = A_0,
\label{bc}
\end{eqnarray}
where $m_0 \alt$ 1 TeV.
We take $A_0 = 0$ in our calculations since the value of $A_0$ 
does not significantly affect the results. 
Most of our conclusions depend only on 
the existence of a soft mass hierarchy and so should be generic,
for example, they should apply to the models in Refs. 
\cite{Dvali:1996rj,Bagger:1999ty}.

At the weak scale, we choose $\tan\beta$ and sign of the $\mu$ parameter
as free parameters.
The value of $|\mu|$ and the Higgs-sector soft breaking bilinear
parameter ($B$) are obtained
by imposing the electroweak symmetry breaking conditions.
In Table I, we present masses in both the ISM and the mSUGRA models 
with $\mu > 0$, $m_0 = 150$ GeV, $m_{1/2} = 200$ GeV, $\tan\beta =$ 3 and 35, 
for the charged Higgs boson ($H^\pm$), 
the lighter chargino ($\chi^\pm_1$) and neutralinos ($\chi^0_{1,2}$),
the lighter top and bottom squark ($\tilde{t}_1$,$\tilde{b}_1$),
the lighter tau slepton ($\tilde{\tau}_1$),
and the first generation squarks and sleptons.


\begin{table}[htb]
\begin{center}
\caption[]{
Masses (GeV) of relevant SUSY particles for $\mu >0$.
}

\medskip

\begin{tabular}{ccccc}
Parameters & 
mSUGRA & 
mSUGRA & 
ISM & 
ISM \\
$m_0$             & 150 & 150  & 150  & 150  \\
$m_{1/2}$         & 200 & 200  & 200  & 200  \\
$A_0$             & 0   & 0    & 0    & 0    \\
$\tan\beta$       & 3   & 35   & 3    & 35   \\
\hline
$m_{H^\pm}$          & 405 & 252 & 398 & 248 \\
$m_{\chi^\pm_1}$     & 142 & 148 & 143 & 150  \\
$m_{\chi^0_2}$       & 144 & 149 & 146 & 150  \\
$m_{\chi^0_1}$       &  76 &  80 &  78 &  81  \\
\hline
$m_{\tilde{t}_1}$    & 312 & 333 & 307 & 327  \\
$m_{\tilde{b}_1}$    & 427 & 373 & 420 & 368  \\
$m_{\tilde{\tau}_1}$ & 172 & 116 & 172 & 117  \\
\hline
$m_{\tilde{u}_L}$    & 469 & 468 & 5015 & 5015  \\
$m_{\tilde{u}_R}$    & 455 & 455 & 5013 & 5013  \\
$m_{\tilde{d}_L}$    & 474 & 475 & 5015 & 5015  \\
$m_{\tilde{d}_R}$    & 455 & 455 & 5013 & 5013  \\
$m_{\tilde{e}_L}$ & 212 & 212 & 5002 & 5002  \\
$m_{\tilde{e}_R}$ & 173 & 164 & 5001 & 5001  \\
$m_{\tilde{\nu}_L}$  & 199 & 197 & 5002 & 5001
\end{tabular}
\end{center}
\end{table}

We have employed one-loop renormalization group equations (RGEs)
to evaluate the weak-scale SUSY particle masses and couplings 
with the boundary conditions in Eq. (\ref{bc}) at the unification scale. 
If two-loop RGEs are used, the third-generation sfermion masses 
at the weak scale might become unphysical 
for $\tilde{m}_{1,2} \agt 22$ TeV and $m_0 \alt 4$ TeV \cite{Tachyon}, 
that corresponds to approximately 
$\tilde{m}_{1,2} \agt 6$ TeV and $m_0 \alt 1$ TeV.

In this letter we present interesting phenomena 
in the ISM and the mSUGRA models: 
the trilepton signature at the upgraded Tevatron, 
the branching fraction of $b \to s\gamma$, 
and the relic density of the neutralino dark matter. 
These three processes can be complementary in distinguishing 
the inverted-scalar-mass hierarchy model and the minimal supergravity model.

\section{Trilepton Signature at the Upgraded Tevatron}

Trileptons from the associated production and decays of 
the lighter chargino ($\chi^\pm_1$) 
and the second lightest neutralino ($\chi^0_2$) 
is probably the most promising channel to search for supersymmetric 
particles at the Tevatron Run II ($\sqrt{s} = 2$ TeV) 
\cite{Trilepton1,Trilepton2,Trilepton3,SUGRA}.
The $\chi^\pm_1 \chi^0_2$ associated production 
has a reasonably large cross section and 
the trilepton background from SM processes can be greatly reduced
with suitable cuts.

The associated production of $\chi^\pm_1 \chi^0_2$ 
occurs via quark-antiquark annihilation 
in the $s$-channel through a $W$ boson and 
in the $t$ and $u$-channels through squarks ($\tilde{q}$).
In both the ISM and mSUGRA models with $m_0 \agt 200$ GeV, 
the up and down squarks are much heavier than the gauge bosons 
and the $s$-channel $W$-resonance amplitude dominates. 
In mSUGRA with $m_0 \alt 150$ GeV, the up and down squarks 
are relatively light and a destructive interference between the $W$ boson 
and the squark exchange amplitudes can suppress the cross section 
by as much as $40\%$, compared to the $s$-channel contribution alone.
For $m_0 \alt 150$ and $\tan\beta \agt 20$,
production of $\tilde{\ell}\tilde{\nu}$ and $\tilde{\ell}\tilde{\ell}$
can enhance the mSUGRA trilepton signal \cite{Trilepton2}.

The Yukawa couplings of the bottom quark ($b$) and the tau lepton ($\tau$)
are proportional to $\tan\beta$ and are greatly enhanced
when $\tan\beta$ is large.
In SUSY grand unified theories, the lighter tau slepton ($\tilde{\tau}_1$)
and the lighter bottom squark ($\tilde{b}_1$) can become lighter than 
other SUSY particles for large $\tan\beta$.

In the ISM model, $\chi^\pm_1$ and $\chi^0_2$ decay dominantly 
into final states with (i) $\tau$ leptons for $\tan\beta \sim 3$, 
and (ii) $b$ quarks and $\tau$ leptons for $\tan\beta \sim 35$.
The contributions from $\tau-$leptonic decays enhance
the trilepton signal substantially when soft cuts
on lepton transverse momenta are used \cite{Trilepton2}.
In mSUGRA, 
the leptonic branching fractions of $\chi^\pm_1$ and $\chi^0_2$ 
depend on the values of $\tan\beta$, $m_{1/2}$ and $m_0$: 
(i) for $\tan\beta \sim 3$, $m_{1/2} \sim 200$ GeV and $m_0 \sim 100$ GeV, 
the dominant decays for $\chi^\pm_1$ and $\chi^0_2$ are 
$\chi^\pm_1 \to \tilde{\tau}_1\nu$, 
$\chi^0_2 \to \tilde{\ell}_R \ell$ and $\tilde{\tau}_1\tau$; 
(ii) for $\tan\beta \sim 35$ and $m_0 \alt 150$ GeV
the $\chi^\pm_1$ and $\chi^0_2$ decay dominantly into final states 
with $\tau$ leptons and $b$ squarks; 
(iii) for $3 \alt \tan\beta \alt 40$ and 180 GeV $\alt m_0 \alt$ 400 GeV, 
$\chi^\pm_1$ and $\chi^0_2$ dominantly decay into final states 
with $q\bar{q}'\chi^0_1$.

Figure 1 shows the branching fractions of 
$\chi^0_2 \to \tau\bar{\tau}\chi^0_1, \ell = e, \mu$, and 
$\chi^0_2 \to \ell\bar{\ell}\chi^0_1, \ell = e, \mu$, 
via virtual and real $\tilde{\tau}$ or $\tilde{\ell}$ 
in the ISM and mSUGRA models for $\tan\beta = 3$ and $\tan\beta = 35$. 
The branching fractions of $\chi^\pm_1 \chi^0_2 \to 3\tau +X$ 
in the ISM model resemble those of large $\tan\beta$ in mSUGRA. 


\begin{figure}[htb]
\centering\leavevmode
\epsfxsize=3.6in\epsffile{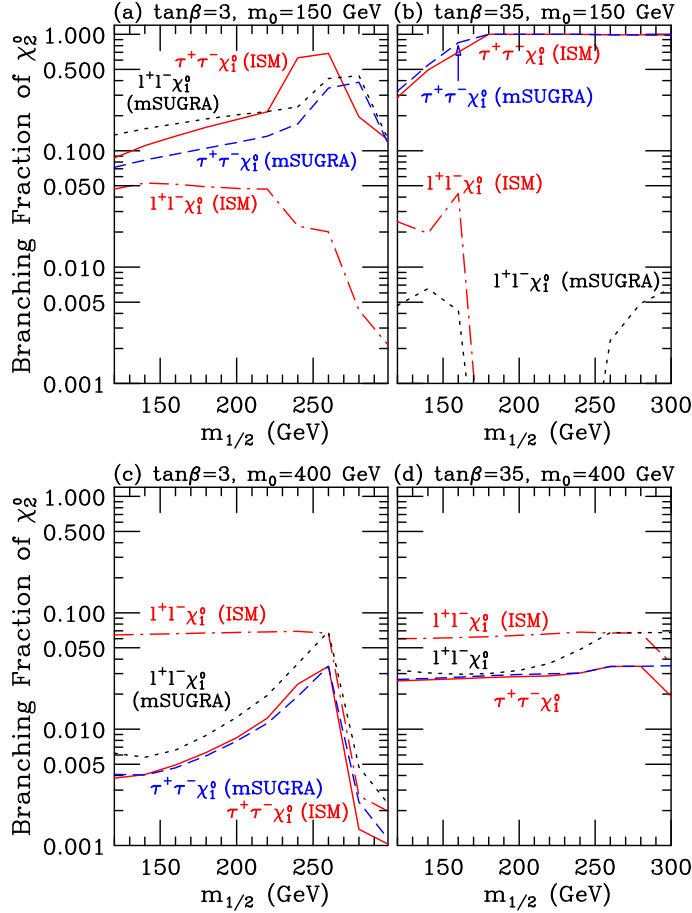}

\smallskip
\caption[]{
Branching fractions of $\chi^0_2 \to \tau\bar{\tau}\chi^0_1$ 
in the ISM (solid) and the mSUGRA (dash) models 
as well as $\chi^0_2 \to \ell\bar{\ell}\chi^0_1$ 
in the ISM (dot-dash) and the mSUGRA (dot) models, 
for $\mu >0$, $\tan\beta = 3$ and 35, and $m_0 = 150$ GeV and 400 GeV.
}
\label{fig:BFZ2}
\end{figure}

To assess the discovery potential of the upgraded Tevatron 
in searching for SUSY particles, we present results from simulations 
for the trilepton signal with an event generator 
and a simple calorimeter including our acceptance cuts.
The ISAJET 7.44 event generator program \cite{ISAJET} 
with the parton distribution functions of CTEQ3L \cite{CTEQ3L} 
is employed to calculate the trilepton signal ($3\ell +\notE_T$) 
from all possible sources of SUSY particles. 
The background from $t\bar{t}$ is calculated with ISAJET as well.

Requiring three isolated leptons in each event 
with $p_T(\ell_{1,2,3}) > 11, 7, 5$ GeV, $|\eta(\ell_{1,2,3})| < 1,2,2$,
along with missing transverse energy $\notE > 25$ GeV, 
and applying other acceptance cuts \cite{Trilepton2},
we find that the major SM backgrounds are
 (i) $q\bar{q}' \to WZ +W\gamma \to \ell\nu \ell\bar{\ell}$
or $\ell'\nu' \ell\bar{\ell}$ ($\ell = e$ or $\mu$)
(ii) $q\bar{q}' \to WZ +W\gamma \to \ell\nu \tau\bar{\tau}$
or $\tau\nu \ell\bar{\ell}$ and subsequent $\tau$ leptonic decays,
with one or both gauge bosons being virtual.
We employed the programs MADGRAPH \cite{Madgraph} 
and HELAS \cite{Helas} to evaluate 
the background cross section of $p\bar{p} \to 3\ell +\notE_T +X$ 
from all these subprocesses.
Additional backgrounds come from production of $t\bar{t}$
and $ZZ \to \ell\bar{\ell}\tau\bar{\tau}$ \cite{Trilepton2,Trilepton3}.
At the upgraded Tevatron with 30 fb$^{-1}$ integrated luminosity (Run III),
we expect about 59 SM background events 
and 38 signal events for a $5 \sigma$ signal \cite{Trilepton2}. 

In Fig. 2, we present the cross section of 
$p\bar{p} \to {\rm SUSY \;\; particles} \to 3\ell +\notE_T +X$, 
versus $m_{1/2}$ for $\mu > 0$, $\tan\beta = 3$ and $\tan\beta = 35$.
The dotted horizontal line denotes the 5$\sigma$ signal cross sections
for ${\cal L} =$ 30 fb$^{-1}$.
At large $\tan\beta$, 
the trilepton cross section in the ISM model 
is slightly larger than that of mSUGRA 
because the first two generation squarks are heavy in the ISM 
and the decays of $\chi^\pm_1\chi^0_2 \to 3\ell +X$ are enhanced.
When $\tan\beta \sim 3$ the trilepton cross section 
is very sensitive to the value of $m_0$:
(i) for $m_0 \sim 100$ GeV, 
the mSUGRA model yields a larger trilepton cross section 
from real decays of 
$\chi^0_2 \to \tilde{\ell}_R \bar{\ell} \to e\bar{e}$; 
(ii) for $m_0 \sim 400$ GeV, the mSUGRA trilepton cross section 
can be smaller by an order of magnitude because 
$\chi^\pm_1$ and $\chi^0_2$ dominantly decay into final states 
with $q\bar{q}'\chi^0_1$.

We conclude that there are major differences 
in the mSUGRA and the ISM predictions for trileptons, 
particularly at low $\tan\beta$, that can differentiate these models.
The mSUGRA predictions show a strong dependence on $m_0$, 
whereas the ISM predictions do not.


\begin{figure}[htb]
\centering\leavevmode
\epsfxsize=3.6in\epsffile{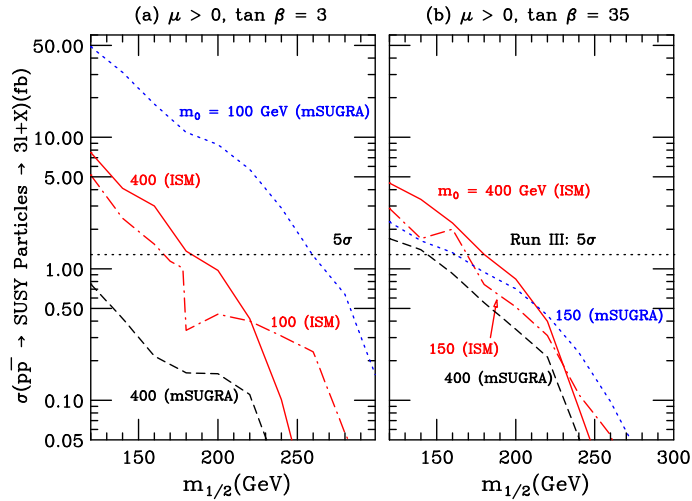}

\smallskip
\caption[]{
Cross section of $p\bar{p} \to {\rm SUSY \; particles} \to 3\ell +X$ 
versus $m_{1/2}$, at $\sqrt{s} = 2$ TeV, with soft acceptance cuts, 
for $\mu >0$, (a) $\tan\beta = 3$ and (b) $\tan\beta = 35$ 
with $m_0 =$ 100 GeV or 150 GeV in the ISM (dot-dash) and the mSUGRA (dot) models 
as well as $m_0 =$ 400 GeV in the ISM (solid) and the mSUGRA (dash) models. 
The horizontal dotted line denotes the cross section 
of a 5 $\sigma$ signal for ${\cal L} =$ 30 fb$^{-1}$.
}
\label{fig:Trileptons}
\end{figure}

In both the mSUGRA and the ISM scenarios the trilepton channel 
could provide valuable information 
about the value of $m_{1/2}$, because the weak-scale gaugino masses 
are related to the universal gaugino mass parameter $m_{1/2}$ by
\begin{eqnarray}
m_{\chi^\pm_1} \sim m_{\chi^0_2} \sim 0.8 m_{1/2}.
\end{eqnarray}
For $m_{1/2} = 120$ GeV and $m_0 = 400$ GeV, 
the chargino mass is 90 GeV for $\tan\beta = 3$ 
and 92 GeV for $\tan\beta = 35$. 
The chargino search at LEP 2 \cite{LEP2} has excluded 
charginos with $m_{\chi^+_1} < 95$ GeV at the 95\% confidence level.
The experiments at the Tevatron may probe a substantial region 
not accessible at LEP 2 \cite{Trilepton2,Trilepton3}.

\section{Constraints from \lowercase{$b \to s \gamma$}}

The experimental measurements of the branching fraction 
$B(b \to s \gamma)$ by the CLEO \cite{CLEO} and LEP collaborations \cite{LEP}
place tight constraints on the parameter space of various models 
and offer guidance for model building.
In the minimal supersymmetric model (MSSM), 
there are dominant contributions from loop diagrams involving 
(i) the $W$ boson and the quarks, 
(ii) the charged-Higgs boson ($H^\pm$) and the quarks, 
and (iii) the charginos ($\chi^\pm_i$) and the squarks. 
The loop diagrams involving neutralinos and gluinos 
are known to be negligible in the MSSM and the mSUGRA model 
\cite{bsg1,bsg2,Dutta:1995fg} and we neglect these contributions here 
in both models. 
As the value of $\tan\beta$ becomes large, 
the form factors from the charged-Higgs-boson diagrams slightly increase, 
while the form factors of the chargino loops are greatly enhanced 
and have opposite sign to the $W-$loop. 
The charged-Higgs-boson loop interferes constructively with the $W$-loop; 
the contributions from chargino loop 
have constructive interference with the $W$-loop when $\mu < 0$, 
but destructive interference with the $W$-loop when $\mu > 0$.
As a result, the predicted value of $B(b \to s \gamma)$ in the MSSM 
is larger than the SM when $\mu < 0$ 
and can be smaller than the SM when $\mu > 0$.
It has been found that the experimental limits of $b \to s\gamma$ disfavor 
most of mSUGRA parameter space
when $\tan\beta \agt 10$ and $\mu < 0$ \cite{bsg2,bsg3}.

Figure 2 shows the branching fraction of $b \to s \gamma$ versus $m_{1/2}$ 
in the ISM and mSUGRA models  
with $\mu > 0$, for (a) $\tan\beta = 3$ and (b) $\tan\beta = 35$. 
Also shown are experimental limits at 95$\%$ confidence level (C.L.) 
($2.0 \times 10^{-4} < B(b \to s \gamma) < 4.5 \times 10^{-4}$) 
from the CLEO collaboration \cite{CLEO}.


\begin{figure}[htb]
\centering\leavevmode
\epsfxsize=4.2in\epsffile{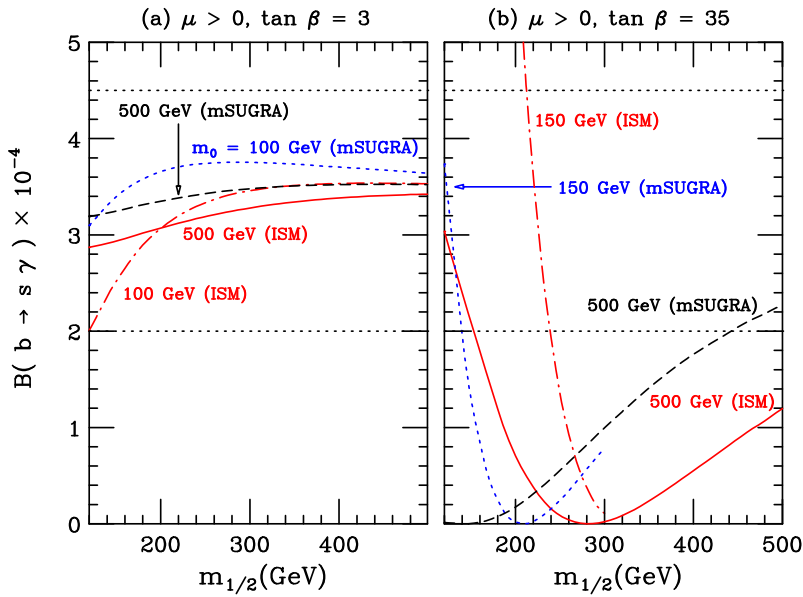}

\smallskip
\caption[]{
Branching fraction of $b \to s \gamma$ versus $m_{1/2}$ 
for $\mu >0$, (a) $\tan\beta = 3$ and (b) $\tan\beta = 35$ 
with $m_0 =$ 100 GeV or 150 GeV in the ISM (dot-dash) and the mSUGRA (dot) models 
as well as $m_0 =$ 500 GeV in the ISM (solid) and the mSUGRA (dash) models. 
The horizontal dotted lines are the 95\% C.L. limits for the CLEO measurements.
}
\label{fig:bsg}
\end{figure}

When $\tan\beta \sim 3$, the predicted branching fractions 
in both the ISM and the mSUGRA models are within the range 
favored by experimental measurements. 
The branching fraction in the ISM model is smaller than that in mSUGRA.

For $\tan\beta \sim 35$ and $m_0 \alt 500$ GeV,
a destructive interference occurs in both the ISM and mSUGRA models 
between loops involving the first two generation squarks 
and loops involving the third generation squarks. 
This destructive interference severely reduces 
the mSUGRA chargino contribution,
and makes $B(b \to s\gamma)$ in mSUGRA much smaller than 
the experimental lower limit.
In contrast, the value of $B(b \to s \gamma)$ in the ISM model 
can be within the allowed experimental limits 
in a sizable region of the $(m_{1/2},m_0)$ plane.

\section{ The Neutralino Relic Density}

In SUSY theories with conservation of $R$-parity
the lightest SUSY particle (LSP) is an attractive candidate
for cosmological dark matter 
because it is stable \cite{SUSYDM}. 
In most of the mSUGRA parameter space, 
the lightest neutralino ($\chi^0_1$) is the LSP. 
For models with gauge mediated SUSY breaking 
or R-parity violation, the $\chi^0_1$ decays 
and cannot be a dark matter candidate.

The matter density of the Universe ($\rho$) is commonly described 
in terms of a relative mass density $\Omega = \rho/\rho_c$ with 
$\rho_c = 3H_0^2/8\pi G_N \simeq 1.88 \times 10^{-29} h^2 \; {\rm g/cm^3}$ 
the critical density to close the Universe. 
Here, $H_0$ is the Hubble constant,
$h = H_0/( 100 \; {\rm km}\; {\rm sec}^{-1}\; {\rm Mpc}^{-1})$, 
and $G_N$ is Newton's gravitational constant.
Based on the matter density $\Omega_m \simeq 0.3\pm 0.05$ inferred from 
cluster X-ray \cite{X-ray} and supernova \cite{Supernovae} observations, 
the baryon density $\Omega_b \simeq 0.019\pm 0.0024$ \cite{Baryon} 
from nucleosynthesis, and 
the Hubble constant $h \simeq 0.65\pm 0.08$ \cite{Hubble} 
the cosmologically interesting region for 
the neutralino dark matter relic density 
($\Omega_{\chi^0_1} = \Omega_m -\Omega_b$) is 
\begin{equation}
0.05 \alt \Omega_{\chi^0_1} h^2 \alt 0.3.
\label{eq:ohs}
\end{equation}

The neutralino dark matter relic density has been studied extensively 
in mSUGRA \cite{SUSYDM,Relic}. 
In Fig. 3, we present the relic density of the neutralino dark matter 
versus $m_{1/2}$ for $\tan\beta = 3$ and $\tan\beta = 35$.
When $\tan\beta \sim 35$, the ISM model and mSUGRA generate 
comparable neutralino relic densities.
For $\tan\beta \sim 3$, the heavy sfermions reduce the 
annihilation cross section of $\chi^0_1\chi^0_1 \to f\bar{f}$ 
and correspondingly increase the neutralino relic density.
Therefore, the value of $\Omega_{\chi^0_1} h^2$ is larger 
in the ISM model with the same parameters.


\begin{figure}[htb]
\centering\leavevmode
\epsfxsize=4.2in\epsffile{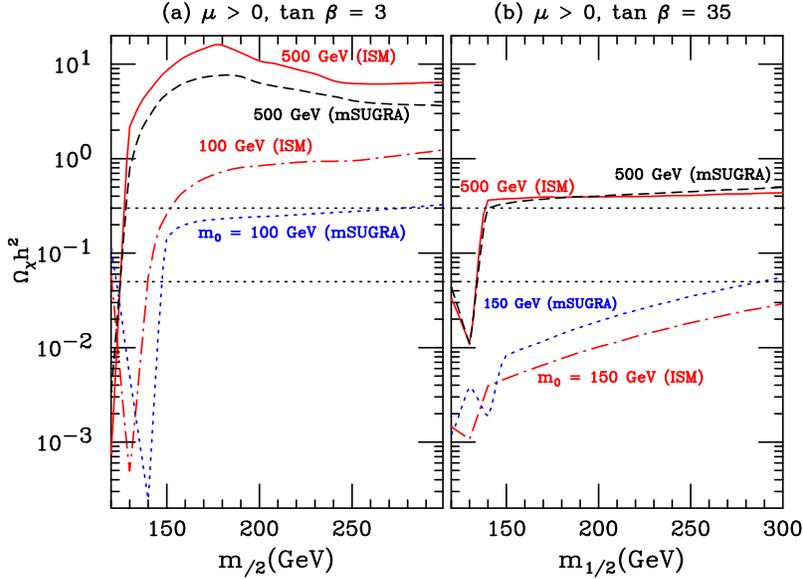}

\smallskip
\caption[]{
$\Omega_{\chi^0_1} h^2$ versus $m_{1/2}$ 
for $\mu >0$, (a) $\tan\beta = 3$ and (b) $\tan\beta = 35$ 
with $m_0 =$ 100 GeV or 150 GeV in the ISM (dot-dash) and the mSUGRA (dot) models 
as well as $m_0 =$ 500 GeV in the ISM (solid) and the mSUGRA (dash) models. 
The horizontal dotted lines denote $\Omega_{\chi^0_1} h^2 = 0.05$ and 0.3.
}
\label{fig:omhs}
\end{figure}

\section{ Conclusions}

The three processes considered in this Letter 
will be complementary in distinguishing 
the inverted-scalar-mass hierarchy model and the minimal supergravity model.
If the nature favors small $\tan\beta \sim 3$, 
searches for the trileptons from $\chi^\pm_1\chi^0_2$ production 
at the upgraded Tevatron can be made 
up to a larger chargino mass in the ISM model for $m_0 \agt 200$ GeV. 
The mSUGRA predictions for trileptons show a strong dependence on $m_0$ 
at low $\tan\beta$, but the ISM predictions do not. 
The mSUGRA model has a larger parameter space 
for a cosmologically interesting relic density 
of the neutralino dark matter.
At high $\tan\beta \agt 35$, 
the experimental constraint from $B(b \to s \gamma)$ allows more parameter 
space in the ISM; the branching fraction in mSUGRA 
is typically too small unless the charginos and the charged Higgs boson 
are very heavy.


\section*{ Acknowledgments}
 
We thank Howie Baer, Bhaskar Dutta and Xerxes Tata for useful discussions. 
This research was supported in part by the U.S. Department of Energy
under Grants No. DE-FG02-95ER40896
and in part by the University of Wisconsin Research Committee
with funds granted by the Wisconsin Alumni Research Foundation.
C.K. thanks Mihoko Nojiri and Darwin Chang for hospitality 
at the Yukawa Institute for Theoretical Physics of the Kyoto University 
and the National Center for Theoretical Sciences 
in the National Tsing-Hua University, 
where part of the research was completed.


%

\end{document}